\documentclass[reprint, superscriptaddress, twocolumn, amsmath, amssymb, aps, prb]{revtex4-2}
\usepackage{graphicx}% Include figure files
\usepackage{dcolumn}% Ali$G_N$ table columns on decimal point
\usepackage{bm}% bold math
\usepackage{placeins}
\usepackage{appendix}
\usepackage{upgreek}

\usepackage{verbatim}
\usepackage[usenames,dvipsnames]{xcolor}
 \usepackage{amsmath}
 \usepackage{amsfonts}
 \usepackage{amssymb}
 \usepackage[colorlinks=true,citecolor=blue,linkcolor=red]{hyperref}

 \usepackage{bbold}     % for \mathbb{1} as a nice unit matrix symbol
 \usepackage[makeroom]{cancel}  % crossed terms in tables
 \usepackage{multirow}    % merge cells vertically in tables
 \usepackage[normalem]{ulem}        % strike-through text
 \usepackage{array}
 \usepackage{pdfpages}
\makeatletter
 \AtBeginDocument{\let\LS@rot\@undefined}
 \makeatother

\begin{document}

%\linenumbers

\title{Probing PbTe-Pb nanowire devices with radio-frequency reflectometry}

\author{Xin-Yi Tang}
\email{equal contribution}
\affiliation{Beijing Academy of Quantum Information Sciences, Beijing 100193, China}
\affiliation{State Key Laboratory of Low Dimensional Quantum Physics, Department of Physics, Tsinghua University, Beijing 100084, China}

\author{Lin Li}
\email{equal contribution}
\affiliation{Beijing Academy of Quantum Information Sciences, Beijing 100193, China}

\author{Zezhou Xia}
\email{equal contribution}
\affiliation{State Key Laboratory of Low Dimensional Quantum Physics, Department of Physics, Tsinghua University, Beijing 100084, China}

\author{Jierong Huo}
\affiliation{College of Semiconductors, Southern University of Science and Technology, Shenzhen 518055, China}

\author{Wenyu Song}
\affiliation{State Key Laboratory of Low Dimensional Quantum Physics, Department of Physics, Tsinghua University, Beijing 100084, China}

\author{Lining Yang}
\affiliation{State Key Laboratory of Low Dimensional Quantum Physics, Department of Physics, Tsinghua University, Beijing 100084, China}

%\author{Weizhao Wang}
%\affiliation{State Key Laboratory of Low Dimensional Quantum Physics, Department of Physics, Tsinghua University, Beijing 100084, China}

%\author{Zeyu Yan}
%\affiliation{State Key Laboratory of Low Dimensional Quantum Physics, Department of Physics, Tsinghua University, Beijing 100084, China}

\author{Zonglin Li}
\affiliation{State Key Laboratory of Low Dimensional Quantum Physics, Department of Physics, Tsinghua University, Beijing 100084, China}

\author{Jiaye Xu}
\affiliation{State Key Laboratory of Low Dimensional Quantum Physics, Department of Physics, Tsinghua University, Beijing 100084, China}

\author{Peilin Li}
\affiliation{State Key Laboratory of Low Dimensional Quantum Physics, Department of Physics, Tsinghua University, Beijing 100084, China}

\author{Runan Shang}
\email{shangrn@baqis.ac.cn}
\affiliation{Beijing Academy of Quantum Information Sciences, Beijing 100193, China}
\affiliation{Hefei National Laboratory, Hefei 230088, China}

\author{Qi-Kun Xue}
\affiliation{Beijing Academy of Quantum Information Sciences, Beijing 100193, China}
\affiliation{State Key Laboratory of Low Dimensional Quantum Physics, Department of Physics, Tsinghua University, Beijing 100084, China}
\affiliation{College of Semiconductors, Southern University of Science and Technology, Shenzhen 518055, China}
\affiliation{Hefei National Laboratory, Hefei 230088, China}
\affiliation{Quantum Science Center of Guangdong-Hong Kong-Macao Greater Bay Area, Shenzhen 518045, China}
\affiliation{Frontier Science Center for Quantum Information, Beijing 100084, China}

\author{Ke He}
\email{kehe@tsinghua.edu.cn}
\affiliation{Beijing Academy of Quantum Information Sciences, Beijing 100193, China}
\affiliation{State Key Laboratory of Low Dimensional Quantum Physics, Department of Physics, Tsinghua University, Beijing 100084, China}
\affiliation{Hefei National Laboratory, Hefei 230088, China}
\affiliation{Frontier Science Center for Quantum Information, Beijing 100084, China}

\author{Hao Zhang}
\email{zhanghao@sustech.edu.cn}
\affiliation{College of Semiconductors, Southern University of Science and Technology, Shenzhen 518055, China}
\affiliation{Quantum Science Center of Guangdong-Hong Kong-Macao Greater Bay Area, Shenzhen 518045, China}

%\date{\today}

\begin{abstract}

We report the implementation of radio-frequency (rf) reflectometry on selective-area-grown PbTe-Pb nanowire devices on a CdTe substrate.  These nanowires are predicted to host Majorana zero modes. We demonstrate the compatibility of the rf technique, including both resistive and capacitive sensing, with these nanowires. The effect of dielectric loss from the CdTe substrate is quantitatively characterized. Furthermore, the feasibility of rf reflectometry is verified under finite magnetic fields where zero-energy modes can emerge. Our results establish the fast control of PbTe quantum devices, paving the way for their applications in topological quantum computation.

\end{abstract}

\maketitle  

\section{Introduction}

Selective-area-grown (SAG) PbTe nanowires have recently attracted much interest because of their potential applications in Majorana zero modes and topological quantum computation \cite{CaoZhanPbTe, Jiangyuying, Erik_PbTe_SAG, Lutchyn2010, Oreg2010, RMP_TQC}. After years of optimization in material growth and device fabrication \cite{PbTe_AB, Fabrizio_PbTe, Zitong_JJ, Wenyu_QPC, Yichun_gap, Yuhao_QPC, Ruidong_PJJ, Vlad_PbTe, PbTe_In, Yichun_SQUID, Zonglin_Anisotropy, Yuhao_dot}, PbTe and PbTe-Pb nanowires can now exhibit transport properties superior to those of InAs and InSb nanowires. For example, high-quality ballistic transport in long-channel PbTe nanowires and quantized Andreev plateaus in PbTe-Pb nanowires have been achieved, indicating an unprecedentedly low disorder level \cite{Yuhao_degeneracy, Wenyu_Disorder, Quantized_Andreev}. Robust zero-energy modes consistent with Majoranas have also been reported \cite{Zhangshan_ZBP, Ruidong_ZBP}. These signatures, however, are still insufficient as conclusive evidence. Definitive confirmation would require the realization of topological qubits and the demonstration of braiding operations. A key requirement for braiding is fast readout of the qubit state \cite{2017_Box_qubit, 2017_PRB_Scalable}. Radio-frequency (rf) reflectometry can meet this requirement. This technique has been implemented in III-V semiconductor devices \cite{rf_2007, rf_2012, rf_2016, rf_2019, rf_2020_NP, rf_2022_PRA, rf_review}. 

Here we report the feasibility of rf reflectometry in SAG PbTe devices. Compared with conventional InAs and InSb devices, the main uncertainties arise from the CdTe substrate and the large dielectric constant of PbTe ($\sim$ 1350). We overcome and quantify these uncertainties and demonstrate rf readout of PbTe-Pb and PbTe nanowire devices. We also show that this technique is compatible with magnetic fields in the range where zero-energy modes can emerge in this material system. Our results open the door to rf control of PbTe quantum devices, representing a step toward PbTe-based topological quantum computation.

\section{Experimental setup}

\begin{figure*}[htb]
\includegraphics[width=\textwidth]{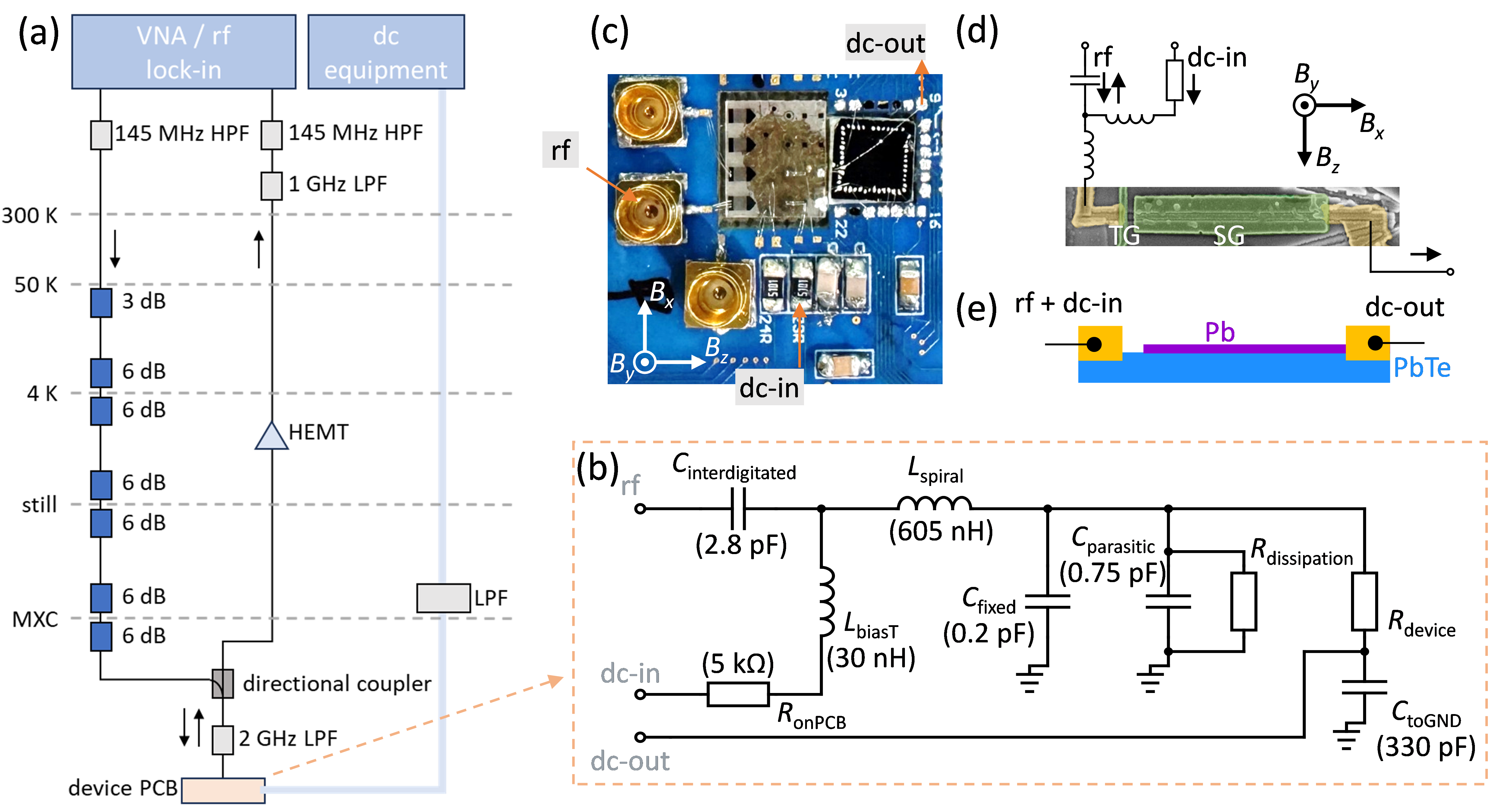}
\centering
\caption{rf-readout system for PbTe nanowires. (a) Schematic of fridge wiring. (b) Circuit diagram of the device PCB. The rf signal is injected from the ‘rf’ port. An off-chip bias-tee design enables simultaneous transport measurement through ‘dc-in’ and ‘dc-out’ ports. (c) An optical image of the device PCB, including the device chip (upper right) and a sapphire substrate carrying superconducting spiral inductors and bias-tees (the middle void). (d) False-colored SEM of the PbTe-Pb NS device. TG and SG refer to tunnel gate and super gate, respectively. (e) Simplified schematic of the device and circuit. For clarity, CdTe substrate/capping layer, PbEuTe buffer/spacing layer, gate oxide/electrodes are not drawn.}

\end{figure*}

Figure 1(a) shows the fridge wiring for rf reflectometry.  The rf signal was generated by a vector network analyzer (VNA) or an rf lock-in amplifier (upper left). After passing through a high-pass filter (HPF) with a cutoff frequency of 145 MHz, the signal was fed into the dilution refrigerator via a coaxial cable. At each temperature stage in the fridge, the signal was attenuated for thermal anchoring. At the mixing-chamber stage, the rf signal was fed into the device PCB through a directional coupler and a low-pass filter (LPF) with a cutoff frequency of 2 GHz. The coupler provides 15 dB attenuation. The total attenuation is $\sim$ 54 dB, which effectively filters out hot photons generated at room temperature.

A photograph of the device printed circuit board (PCB) is shown in Fig. 1(c). Its equivalent circuit diagram is  sketched in Fig. 1(b). For a complete view, we refer to Fig. 6 in the Appendix. The rf signal enters a Miniature SubMiniature Push-On (Mini SMP) connector on the PCB, marked by the left orange arrow in Fig. 1(c). The center pin of the Mini SMP connector is further connected to a large capacitor ($C_{\text{interdigitated}}$ = 2.8 pF) through wire bonding. Meanwhile, a dc voltage signal is added to the rf signal after passing through a 5 k$\Omega$ resistor ($R_{\text{onPCB}}$, lower arrow in Fig. 1(c)) and a 30 nH spiral inductor ($L_{\text{biasT}}$). $C_{\text{interdigitated}}$ and $L_{\text{biasT}}$ together form a bias tee that prevents interference between the rf and dc channels. The interdigitated capacitor and spiral inductor were both fabricated on a sapphire substrate by etching a 100-nm-thick Nb film (see Appendix for details).  The sapphire substrate was mounted/glued in the center void of the PCB, next to the device chip.

The rf and dc signals pass through an inductor ($L_{\text{spiral}}$), and are connected to the source contact of the PbTe-Pb device through wire bonding. A ceramic capacitor on the PCB ($C_{\text{fixed}}$ = 0.2 pF) is connected in parallel with the device.  The device drain is connected to another ceramic capacitor ($C_{\text{toGND}}$ = 330 pF) and to dc-out for current measurement using a preamplifier.

The scanning electron micrograph (SEM) of the device and a simplified circuit diagram are shown in Figs. 1(d-e).  A dc bias voltage ($V$) is applied to the device source, and the current $I$ is measured at the drain. Meanwhile, the rf signal is sent to the LCR circuit through the rf port in Fig. 1(b). The reflected signal is directed to another coaxial line through the directional coupler. This signal is then amplified by roughly 30 dB using a high-electron-mobility transistor (HEMT) at 4 K and measured at room temperature using the VNA or rf lock-in. From the perspective of the PCB, the reflected signal $S_{11}$ is measured. From the fridge circuiting, it corresponds to the transmitted signal $S_{21}$ between two coaxial ports. In the following, we refer to it as $S_{21}$.

The PbTe-Pb device has two top gates (TG and SG). Details of its growth and fabrication can be found in the Appendix and in our previous works \cite{Wenyu_Disorder, Zhangshan_ZBP}. $V_{\text{TG}}$ tunes the tunnel barrier, while $V_{\text{SG}}$ tunes both the proximitized region and, through cross talk, the barrier region. In the circuit diagram in Fig. 1(b), the device is represented as a gate-tunable resistor ($R_{\text{device}}$). In parallel, we also include a parasitic capacitor $C_{\text{parasitic}}$ and a fixed resistor $R_{\text{dissipation}}$. The former originates from the PCB, bonding pads, sample wiring, and the device, whereas the latter accounts for dielectric loss in the device chip. $C_{\text{parasitic}}$ = 0.75 pF was extracted based on independent characterization. The value of $R_{\text{dissipation}}$ can be inferred from our simulation and will be discussed later.  Note that $R_{\text{dissipation}}$ is introduced only in the rf frequency range to account for dielectric loss; it approaches $\infty$ for dc measurements.

We denote the impedance of the circuit in Fig. 1(b) at the rf port (Mini SMP) as $Z_l (\omega)$, where $\omega = 2\pi f$ is the angular frequency of the signal. The impedance of the coaxial cable is $Z_0$ = 50 $\Omega$. From the telegraph equation \cite{rf_review}, the reflection coefficient $\Gamma$, i.e., the ratio between the reflected signal voltage and the incident signal voltage, can be expressed as
\begin{equation}
\begin{aligned}
 \Gamma (\omega) = \frac{Z_l(\omega)-Z_0}{Z_l(\omega)+Z_0}.
\end{aligned}
\end{equation}
In the experiment, $|S_{21}|(\omega) = 20\log_{10}{|\Gamma(\omega)|}$ (dB) was measured. To analyze the circuit behavior near its resonance frequency ($\omega_0$), we make a rough approximation by neglecting the the bias-tee section ($C_{\text{interdigitated}}$ and $L_{\text{biasT}}$), because $\omega_0$ is far from the resonance of the bias tee. $C_{\text{toGND}}$ can also be neglected because it is shorted by the dc-out line. We can then derive
\begin{equation}
\begin{aligned}
Z_l(\omega) \approx \text{i}\omega L + (\text{i}\omega C+\frac{1}{R})^{-1},
\end{aligned}
\end{equation}
 where $L=L_{\text{spiral}}$, $C=C_{\text{fixed}}+C_{\text{parasitic}}$, $R=R_{\text{dissipation}} || R_{\text{device}}=(R_{\text{dissipation}}R_{\text{device}})/(R_{\text{dissipation}}+R_{\text{device}})$. Near the cavity resonance $\omega_0=1/\sqrt{LC}$, Eq. [2] can be expanded using $\omega=\omega_0+\delta\omega$. Omitting small terms gives $Z_l(\omega) \approx L/(RC)$, provided that $L/(CR^2)\ll 1$. Therefore, near the circuit resonance, the gate-tunable $R_{\text{device}}$ sequentially affects $R$, $Z_l$, and $\Gamma$. This change in device resistance is readout as a change of the reflected signal power $|S_{21}|$. We note that the analysis above is based on rough approximations. For more accurate analysis, we refer to simulations discussed below that include all elements in Fig. 1(b).

\section{rf and dc probing of a P\MakeLowercase{b}T\MakeLowercase{e} -P\MakeLowercase{b} device}

Figure 2(a) shows the rf response of the NS device measured using a VNA at a magnetic field of 0 T. The bias voltage $V$ was set to $-2$ mV, i.e., outside the superconducting gap. $V_{\text{TG}}$ was kept fixed at $-5$ V throughout the measurement, and $V_{\text{SG}}$ was scanned to tune the barrier region via cross talk. At $V_{\text{SG}} = -4$ V, the device is pinched off, and a dip appears in the frequency $f$ scan. The black curve in Fig. 2(b) is its line cut. This dip corresponds to the resonance of the LCR circuit, with a resonance frequency of 241.1 MHz. As $V_{\text{SG}}$ becomes more positive, the dip becomes smaller and eventually disappears. Figure 2(b) shows this evolution with several line cuts.

This rf evolution arises from the gate-tunable $R_{\text{device}}$ in the LCR circuit. To map the correspondence between the reflected rf signal voltage ($V_{\text{rf}}$) and the device conductance $G\equiv \text{d}I/\text{d}V$, i.e., $1/R_{\text{device}}$, we fixed $f$ at 241.1 MHz and measured both $G$ and $V_{\text{rf}}$ simultaneously. $V_{\text{rf}}$ was measured using an rf lock-in, while $G$ was measured using a standard two-terminal setup. $R_{\text{onPCB}}$ and fridge filters were subtracted and excluded from $G$. Figure 2(c) shows their evolution as a function of $V_{\text{SG}}$. Based on this correspondence, we labeled the conductance values for the line cuts in Fig. 2(b). Figure 2(d) plots $V_{\text{rf}}$ versus $G$. The red curve is a polynomial fit between $\ln G$ and $V_{\text{rf}}$. Details of the fitting can be found in the Appendix.

\begin{figure}[htb]
\includegraphics[width=\columnwidth]{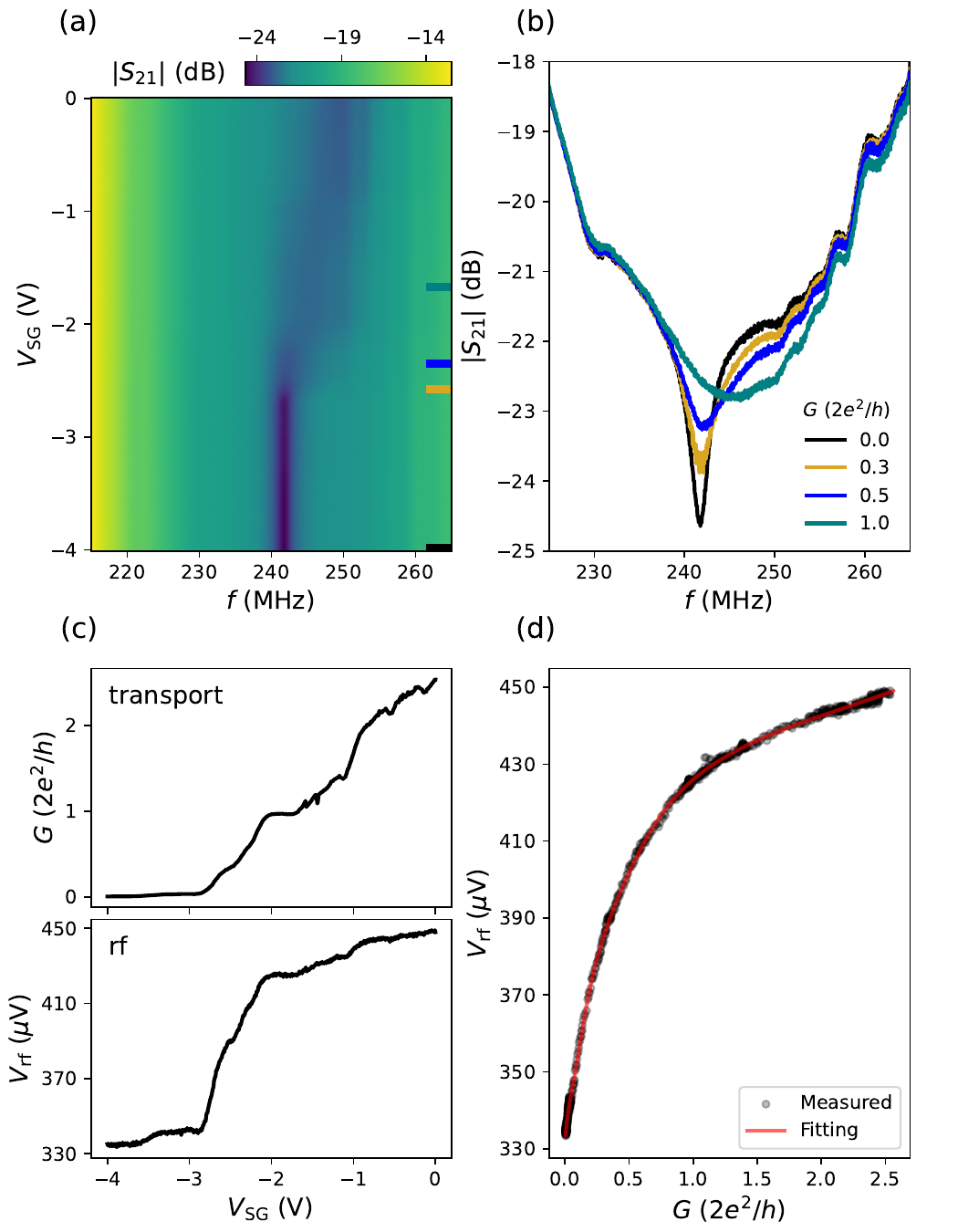}
\centering
\caption{Gate-tunable reflectometry. (a) $|S_{21}|$ as a function of $V_{\text{SG}}$ and the rf frequency $f$. (b) Line cuts at several gate voltages (marked in (a)). (c) Transport measurement (upper) and the rf signal (lower). $f$ is fixed at 241.1 MHz. (d) $V_{\text{rf}}$ vs $G$. The red curve is the fitting.  }
\end{figure}

\begin{figure}[htb]
\includegraphics[width=\columnwidth]{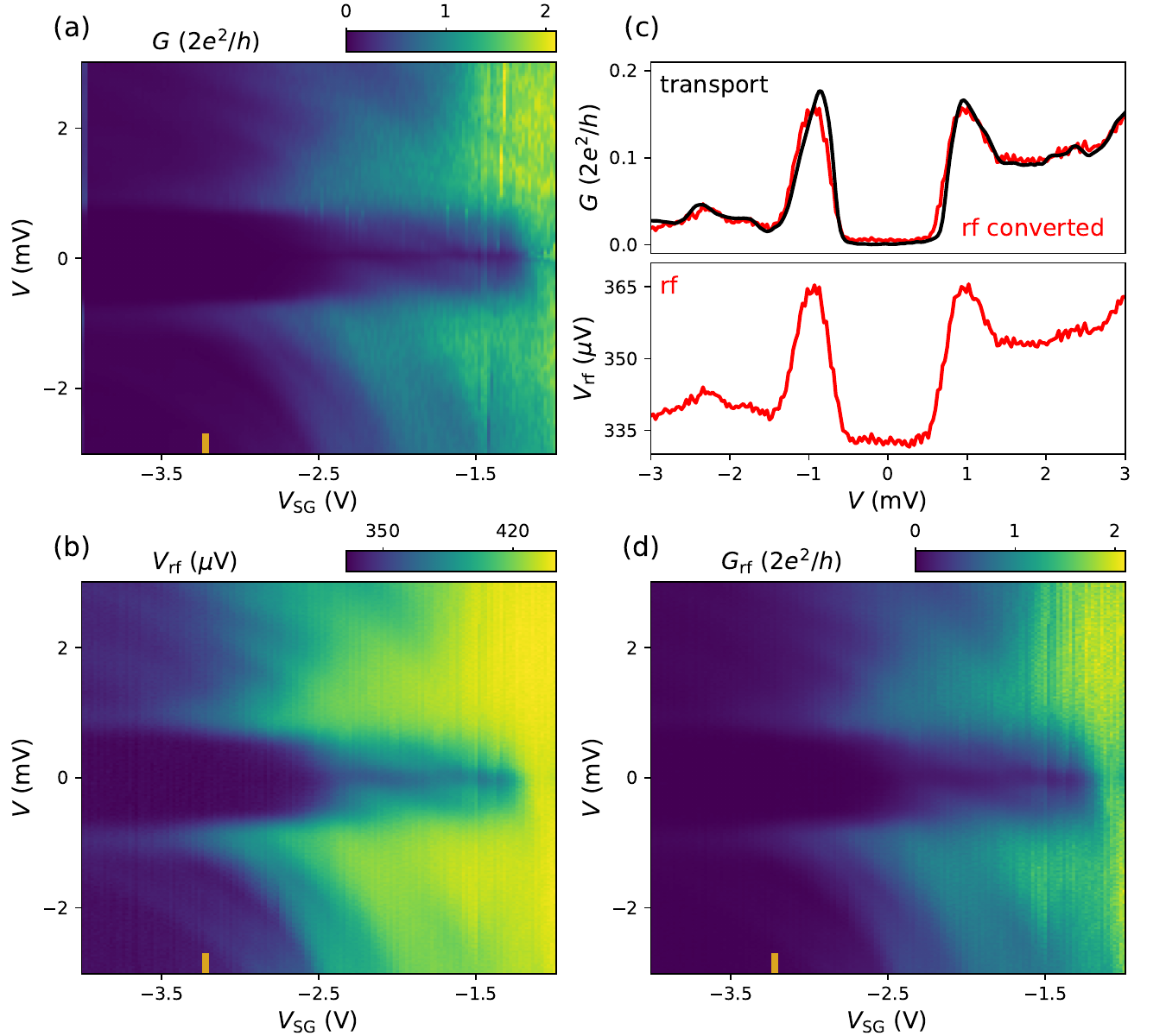}
\centering
\caption{Comparison of rf and dc probing of the PbTe-Pb nanowire. (a) dc conductance map as a function of $V$ and $V_{\text{SG}}$. (b) rf measurement under the same gate and bias setting. (c) Line cuts from (a) (upper, black) and (b) (lower) at $V_{\text{SG}}$ = $-3.22$ V (marked as orange bars). The red curve in the upper panel is the rf signal converted into $G$. (d) Conductance ($G_{\text{rf}}$) map converted from the rf signal in (b).   }
\end{figure}

The maximum depth of the cavity dip (black curve in Fig. 2(b)) is $\sim$ 3 dB. Although this value is sufficient for our current requirement, it is small compared with the resonance signal ($\sim$ 20 dB) in rf circuits based on InAs nanowires \cite{rf_2019}. We attribute the small dip to impedance mismatch caused by $C_{\text{parasitic}}$ and $R_{\text{dissipation}}$. Unlike other circuit elements with predefined parameters, these two terms were unknown before the measurement. $R_{\text{dissipation}}$ can originate from parasitic losses in the CdTe substrate, the dielectric mask (Si$_3$N$_4$), and the Ag paste used during sample wiring. To estimate their values, we simulated the circuit in Fig. 1(b) using a software Advanced Design System. Based on the simulation (shown in Appendix Fig. 7), we find that $R_{\text{dissipation}} =$  4 k$\Omega$ agrees qualitatively well with our results in terms of the dip variation for $G$ between 0 and $2e^2/h$. $L_{\text{spiral}} =$ 605 nH also differs from our designed value of 389 nH because the inductance of the bonding wire that connects the spiral inductor to the device is included in $L_{\text{spiral}}$. This contribution can depend on the sample and wiring. The simulated circuit impedance at the resonance frequency is $Z_l = 125.2 - 5.9\text{i}$ ($\Omega$), assuming that the device is pinched off. This impedance is indeed far from 50 $\Omega$, the impedance of the coaxial cable. The rough estimate using $L/(RC)$ gives 159 $\Omega$, close to the simulation but with a small deviation, likely due to the omission of the segment of bias tee. As the device is turned on, $R$ decreases and $Z_l$ moves farther away from 50 $\Omega$, resulting in the smaller dip shown in Fig. 2(b). In Fig. 5, we show circuits with better impedance matching, which results in a deeper resonance.

We next compare this rf method with dc transport measurements at zero field. Figure 3(a) shows the standard conductance map of the device: $G$ versus $V$ and $V_{\text{SG}}$. A clear superconducting gap can be resolved near $\pm$ 0.9 mV. This two-dimensional (2D) map has 201 points in bias $V$ and 151 points in $V_{\text{SG}}$. The data-acquisition time is approximately 400 min. The frequency and time constant used in the low-frequency lock-in amplifier were set to 33.7 Hz and 300 ms, respectively. By comparison, a similar 2D map obtained using the rf method (Fig. 3(b)) requires only 39 min, an order of magnitude faster than the traditional dc measurement. The ranges and numbers of data points in gate and bias in Fig. 3(b) are identical to those in Fig. 3(a). The rf frequency was fixed at 241.1 MHz. The bandwidth and number of averages were chosen to be  8.872 kHz  and 1000, respectively. Instead of point-by-point data acquisition in the dc measurement scheme, the rf scan sends a triangular-wave-like bias signal into the device and records the reflected signal altogether before stepping $V_{\text{SG}}$ to the next value.

For comparison, the upper and lower panels of Fig. 3(c) show line cuts from Fig. 3(a) and Fig. 3(b), respectively, at $V_{\text{SG}} = -3.22$ V. Based on the mapping established in Fig. 2(d) (red curve), we convert the rf signal ($V_{\text{rf}}$) into conductance ($G_{\text{rf}}$) and plot it as the red curve in the upper panel of Fig. 3(c). The transport and rf measurements agree reasonably well. For completeness, Fig. 3(d) shows the full conductance map ($G_{\text{rf}}$) converted from Fig. 3(b). The color bar in Fig. 3(d) is identical to that in Fig. 3(a). We note that $V$ is the total bias drop across the device, fridge filters, and $R_{\text{onPCB}}$. The order-of-magnitude reduction in data-acquisition time offers a significant advantage by accelerating the exploration of PbTe-Pb devices for Majorana searches.

\section{Magnetic field compatibility}

We next verify the feasibility of this technique in a magnetic field ($B$), especially in the field range where Majorana zero modes may occur. We have reported possible Majorana signatures in PbTe-Pb nanowires \cite{Zhangshan_ZBP, Ruidong_ZBP}. The optimal $B$ is near 0.2 T, and its direction is nearly out of plane, i.e., $B_y$ (see Figs. 1(c-d) for the coordinate axes). Figure 4(a) shows the resonator response with $B$ applied along this direction. The NS device is pinched off, i.e., $V_{\text{SG}} = -4$ V. As $ B_y $ increases, the resonance shifts toward lower frequency. We attribute this shift to the increase in kinetic inductance of the superconducting spiral inductors, as the magnetic field reduces the Cooper-pair density. 

\begin{figure}[htb]
\includegraphics[width=\columnwidth]{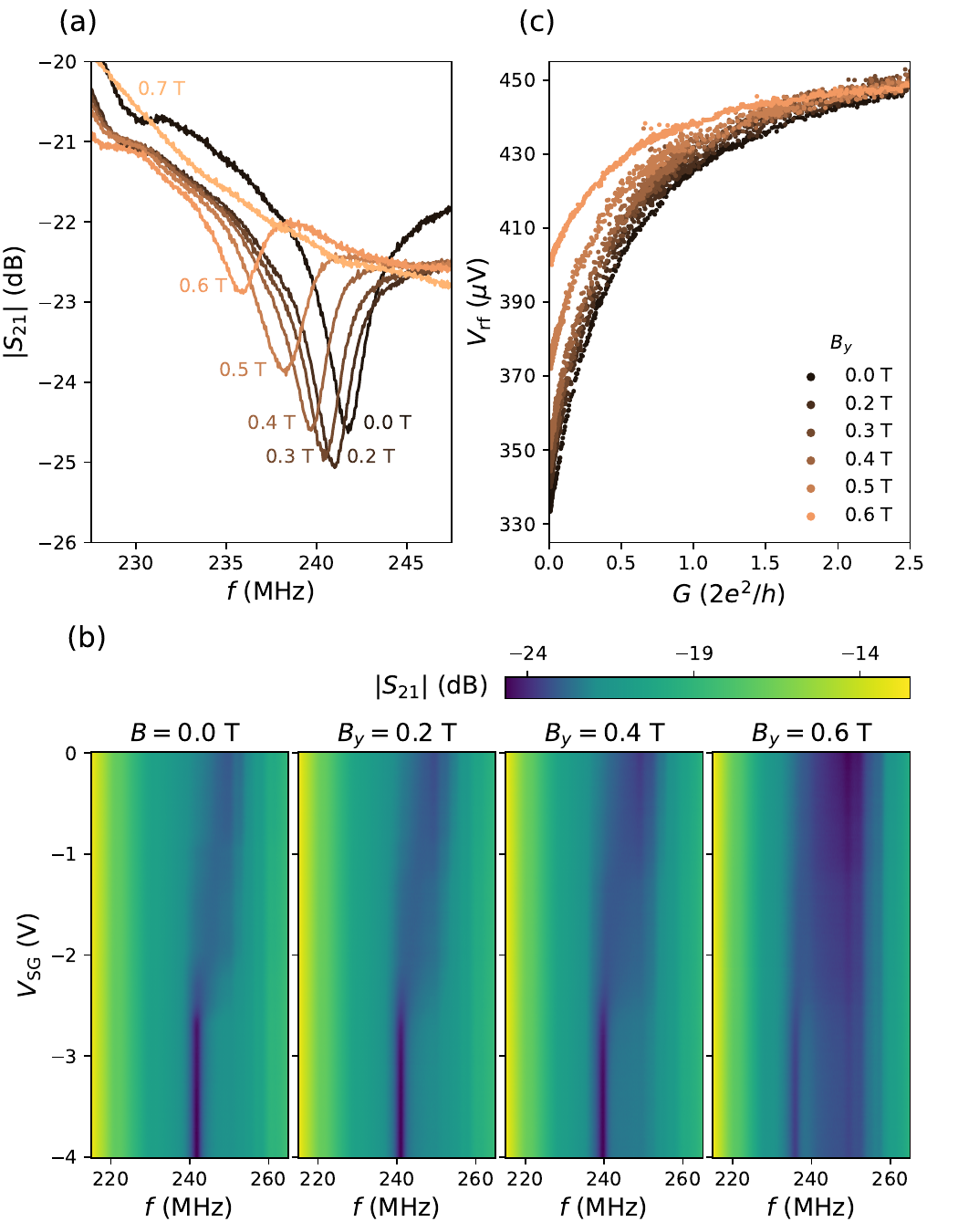}
\centering
\caption{Magnetic field compatibility. (a) The resonator response at different $B_y$'s (out-of-plane). $V_{\text{SG}}$ is fixed at $-4.0$ V. (b) The reflectometry signal as a function $V_{\text{SG}}$ at $B_y$ of 0.0, 0.2, 0.4, 0.6 T, respectively. (c) rf signal ($V_{\text{rf}}$) vs $G$ obtained at different fields. }
\label{fig1}
\end{figure}

\begin{figure*}[htb]
\includegraphics[width=\textwidth]{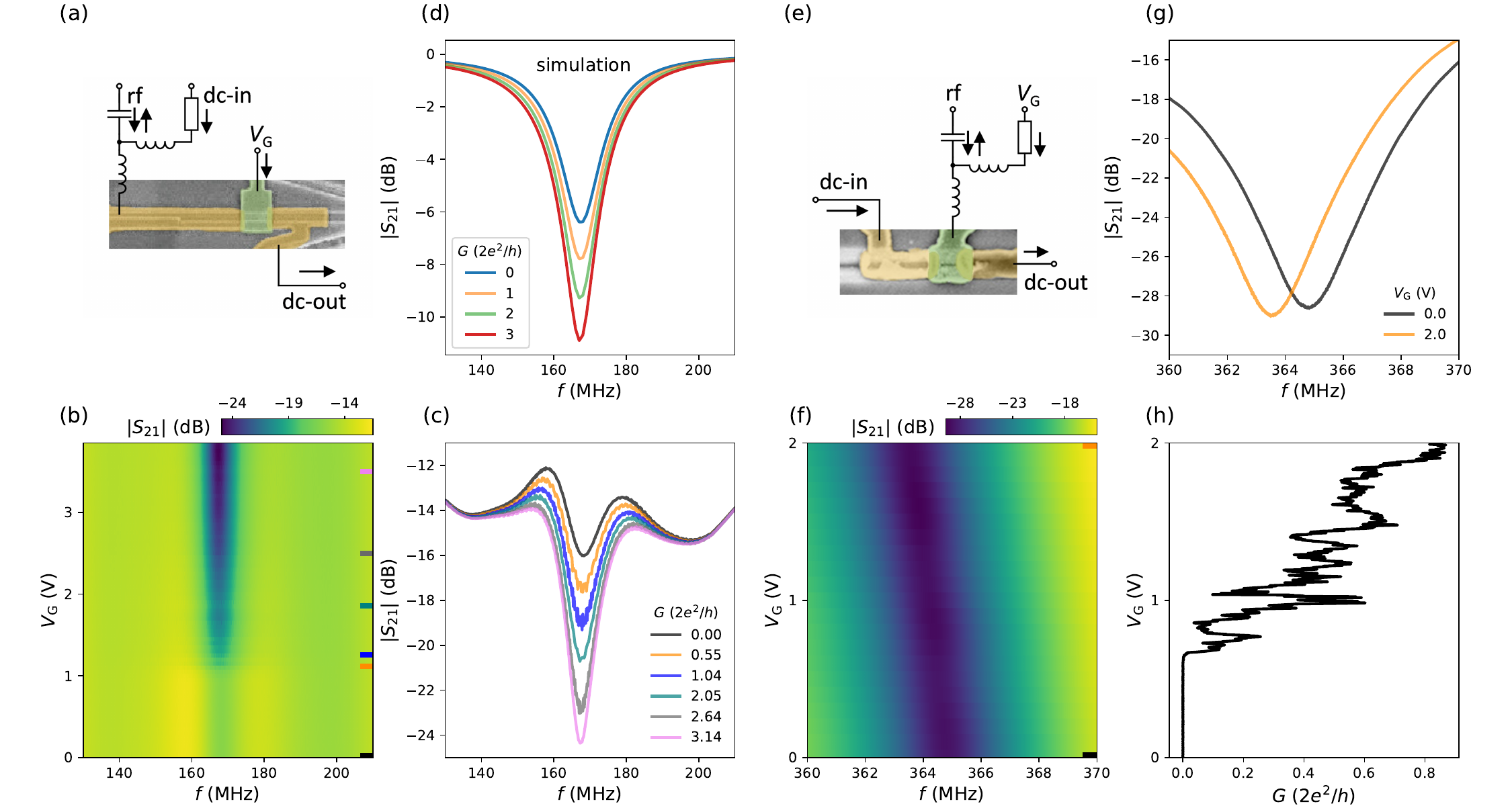}
\centering
\caption{Resistive and capacitive sensing of PbTe devices. (a) SEM of a PbTe device using the resistive readout. (b) $|S_{21}|$ as a function of $V_{\text{G}}$ and the rf frequency. (c) Line cuts at several gate voltages marked in (b). (d) Simulated cavity response for different device conductance. (e) SEM of a PbTe device using capacitive readout. (f) $|S_{21}|$ as a function of $V_{\text{G}}$ and the rf frequency. (g) Two line cuts from (f) at $V_{\text{G}}$ = 0 V and 2 V, respectively. (h) Conductance of this device as a  function of $V_{\text{G}}$. }
\end{figure*}

The cavity resonance is fully suppressed at $B_y$ = 0.7 T. Notably, up to 0.3 T, the shape of the resonance dip remains nearly unchanged. Figure 4(b) shows the gate response of the resonance at different fields. $V$ was fixed at $-2$ mV. At $B_y =$ 0.2 T, the resonance is nearly identical to that at 0.0 T. Next, we fixed the frequency to the resonance for each field and measured the rf response $V_{\text{rf}}$ and conductance $G$ simultaneously. Figure 4(c) shows their relation at each field. To facilitate comparison, we add a vertical offset to each curve so that all curves are aligned at $G= 2.5\times 2e^2/h$. The mapping barely changes for fields below 0.4 T. This field exceeds the range of robust zero modes (near 0.2 T), demonstrating the compatibility of our rf technique with Majorana searches. For fields above 0.4 T, the $V_{\text{rf}}$ range starts to shrink because of suppression of superconductivity of the spiral inductor and resonator degradation.

For completeness, we also tested the resonator response for in-plane fields, i.e., $B_z$ and $B_x$, as shown in Appendix Fig. 8. For these directions, the resonator can sustain fields up to 1 T before degradation.

\section{Resistive and capacitive sensing of P\MakeLowercase{b}T\MakeLowercase{e}  nanowires}

In addition to the NS device, we implemented this rf technique in PbTe field-effect devices. Figure 5(a) shows the false-color SEM. The PbTe nanowire is contacted by two metallic leads (yellow) and then covered by a thin Al$_2$O$_3$ dielectric layer and a top gate (green). The rf probe is connected to the source, similar to the NS device. Figure 5(b) shows the reflected amplitude $|S_{\text{21}}| $ as a function of the top-gate voltage $V_{\text{G}} $ and the rf frequency $f$. Line cuts at several gate voltages are plotted in Fig. 5(c). Compared with Fig. 2, the resonance here is deeper, reaching 10 dB for $G$ = $3.14\times 2e^2/h$ (see the pink curve in Fig. 5(c)). The sharper resonance is due to improved impedance matching. The PCB design is different from that used for the NS device. The circuit-element parameters are: $C_{\text{fixed}}$ = 0.2 pF, $C_{\text{parasitic}}$ = 4.3 pF, $L_{\text{spiral}}$ = 520 nH, and $R_{\text{dissipation}}$ = 2.5 k$\Omega$; remaining elements are identical to those in Fig. 1(b). We calculate the circuit impedance at the resonance frequency ($f$ = 167 MHz) to be $27.9-2.7\text{i}$ ($\Omega$) for $G$ = $3\times2e^2/h$. Although this impedance is still not matched to 50 $\Omega$, it is much closer than that of the NS device. Since the impedance is smaller than 50 $\Omega$, a decrease in conductance further decreases the impedance. Correspondingly, the dip becomes shallower, exhibiting a trend opposite to that in the $V_{\text{G}}$ scan in Fig. 2(a). Figure 5(d) shows the simulated resonator response for different device conductances using the parameters mentioned above. The simulation qualitatively captures the experimental results in Fig. 5(c).

In resistive sensing, the resonance frequency barely changes, whereas the sharpness of the resonance dip varies as the device conductance is tuned. In Fig. 5(e) for another field-effect device, we connected the rf probe to the top gate instead of the contact. Here in this capacitive sensing geometry, $V_{\text{G}}$ tunes the device capacitance, causing a shift in the resonance frequency. Figure 5(f) shows the gate scan. The resonance frequency shifts by $\sim$ 1.2 MHz when $V_{\text{G}}$ is scanned from 0.0 V to 2.0 V, see the line cuts in Fig. 5(g).

The device capacitance is governed by the gate geometry (area and dielectric thickness) as well as by the electron distribution in the PbTe nanowire. The former is fixed once the device is fabricated, whereas the latter can be affected by $V_{\text{G}}$. A more positive $V_{\text{G}}$ attracts more electrons to the top surface of the nanowire. This accumulation layer shortens the effective spacing between the capacitor plates and increases the capacitance. The resonance frequency ($f_C=1/(2\pi \sqrt{LC})$) therefore decreases, as shown in Fig. 5(f). The device conductance as a function of $V_{\text{G}}$ is shown in Fig. 5(h). The conductance fluctuations are not correlated with the resonance frequencies during the $V_{\text{G}}$ scan. Moreover, the resonance continues to shift toward higher frequencies even after the device is pinched off. We attribute this behavior to gate-induced charges in the CdTe substrate. The conduction band of CdTe is only 0.3 eV above that of PbTe, making it possible that $V_{\text{G}}$ depletes electrons from PbTe into CdTe.

From the total frequency shift $\Delta f_C$ of 1.2 MHz near $f_C$ of 365 MHz, we can calculate the relative capacitance change $|\Delta C/C|=2|\Delta f/f| = 0.66\%$. The estimated capacitance of 590 fF thus yields $\Delta C \sim$3.9 fF. This value is comparable with the capacitance of the PbTe device estimated using a parallel plate model $C_{\text{PbTe}} = \epsilon A/d = 4.8$ fF. The area $A=200\times100$ nm$^2$ and $d=50$ nm. The 3.9 fF shift in Fig. 5(f) suggests that our rf circuit can differentiate a capacitance change of $\sim$1 fF, sufficient for parity readout \cite{MS_2025}. 

\section{Conclusion and outlook}

To summarize, we have implemented the radio-frequency reflectometry in PbTe and PbTe-Pb nanowire devices. For resistive sensing, the data-acquisition time can be shortened by an order of magnitude compared with traditional dc measurements. The effect of the CdTe substrate is quantified by a dissipation resistor. For capacitive sensing, a gate-tunable cavity shift is observed. We further demonstrate the compatibility of this rf setup with magnetic fields, especially in the field range where signatures of Majorana zero modes have been reported. Our results enable radio-frequency experiments in PbTe quantum devices. Future efforts could focus on improved printed circuit board design (e.g., by including in-situ tunable capacitors) for better impedance matching and reduced substrate dissipation.

\section{Acknowledgment}

This work is supported by National Natural Science Foundation of China (92565302) and Quantum Science and Technology-National Science and Technology Major Project (2021ZD0302400). W. S. acknowledges the Postdoctoral Fellowship Program and China Postdoctoral Science Foundation (Grant No. BX20250167 and No. 2025M783399). 

\section{Data availability}

The data that support the findings of this article can be found at \cite{rawdata}.

\appendix

\section{Methods}

\textbf{Spiral inductor fabrication.} A 100-nm-thick Nb film was sputtered on a sapphire substrate. The film was then covered by a photoresist (S1813, 3000 rpm) and baked at 115 $^{\circ}$C for 90 s. Patterns including spiral inductors, bias tees, and bonding pads were defined using direct laser writing. The chip was developed in MF319 for 45 s, followed by reactive ion etching (SF$_6$, 30 sccm Ar, 10 sccm, 100 s). The residual resist was removed in acetone.

\textbf{Nanowire growth and device fabrication.} Three PbTe nanowire devices were studied, including one PbTe-Pb NS device and two PbTe field-effect devices. The nanowires were grown selectively on a CdTe substrate. The PbTe-Pb nanowire consists of 8 nm $\text{Pb}_{0.99}\text{Eu}_{0.01}\text{Te}$ buffer,  30 nm $\text{PbTe}$,  5 nm $\text{Pb}_{0.97}\text{Eu}_{0.03}\text{Te}$ spacing,  7 nm $\text{Pb}$ and 11 nm $\text{CdTe}$ capping. The PbTe nanowire used for resistive sensing consists of 5 nm $\text{Pb}_{0.97}\text{Eu}_{0.03}\text{Te}$ buffer,  12.5 nm $\text{PbTe}$, and 5 nm $\text{Pb}_{0.95}\text{Eu}_{0.05}\text{Te}$ capping. The PbTe nanowire used in capacitive sensing consists of 5 nm $\text{Pb}_{0.99}\text{Eu}_{0.01}\text{Te}$ buffer,  12 nm $\text{PbTe}$,  5 nm $\text{Pb}_{0.99}\text{Eu}_{0.01}\text{Te}$ and 11 nm $\text{CdTe}$ capping. Details of the growth can be found in Ref. \cite{Wenyu_Disorder}. 

The fabrication of PbTe nanowire devices consists of three electron-beam lithography steps. First, Ti/Au metallic contacts were deposited. Prior to contact deposition, in situ Ar plasma etching was performed to remove surface oxides and the PbEuTe/CdTe capping layer. An Al$_2$O$_3$ dielectric layer was then deposited by atomic layer deposition. Finally, top gates were deposited by Ti/Au evaporation. For the PbTe-Pb device, before these steps, most of the Pb film on the substrate was etched using Ar ion milling to prevent short circuiting.

\textbf{Fitting.} In Fig. 2(d), we assume a polynomial relationship between $\ln G$ and $V_{\text{rf}}$: 
\begin{equation}\begin{aligned}
\ln G =&a_6 \cdot (V_{\text{rf}})^6+  a_5\cdot (V_{\text{rf}})^5 + a_4\cdot (V_{\text{rf}})^4 +a_3\cdot (V_{\text{rf}})^3\\
& +a_2\cdot (V_{\text{rf}})^2 +a_1\cdot (V_{\text{rf}}) +a_0,
\end{aligned}\end{equation}
as this guarantees $G>$ 0. We find that a polynomial expression up to $(V_{\text{rf}})^6$ can give precise and satisfactory fit shown as the red curve in Fig. 2(d). Here, $V_{\text{rf}}$ is in units of $\upmu$V, $a_6 = -5.58587503\times{10^{-11}}$, $a_5 = 1.34033713\times{10^{-7}}$, $a_4 = -1.33805403\times{10^{-4}}$, $a_3 = 7.11366689\times{10^{-2}}$, $a_2 = -2.12427991\times{10^{1}}$, $a_1=3.37849600\times{10^{3}}$, $a_0 = -2.23587555\times{10^{5}}$.

\section{Details of the PCB circuit}

\begin{figure*}[htb]
\includegraphics[width=0.95\textwidth]{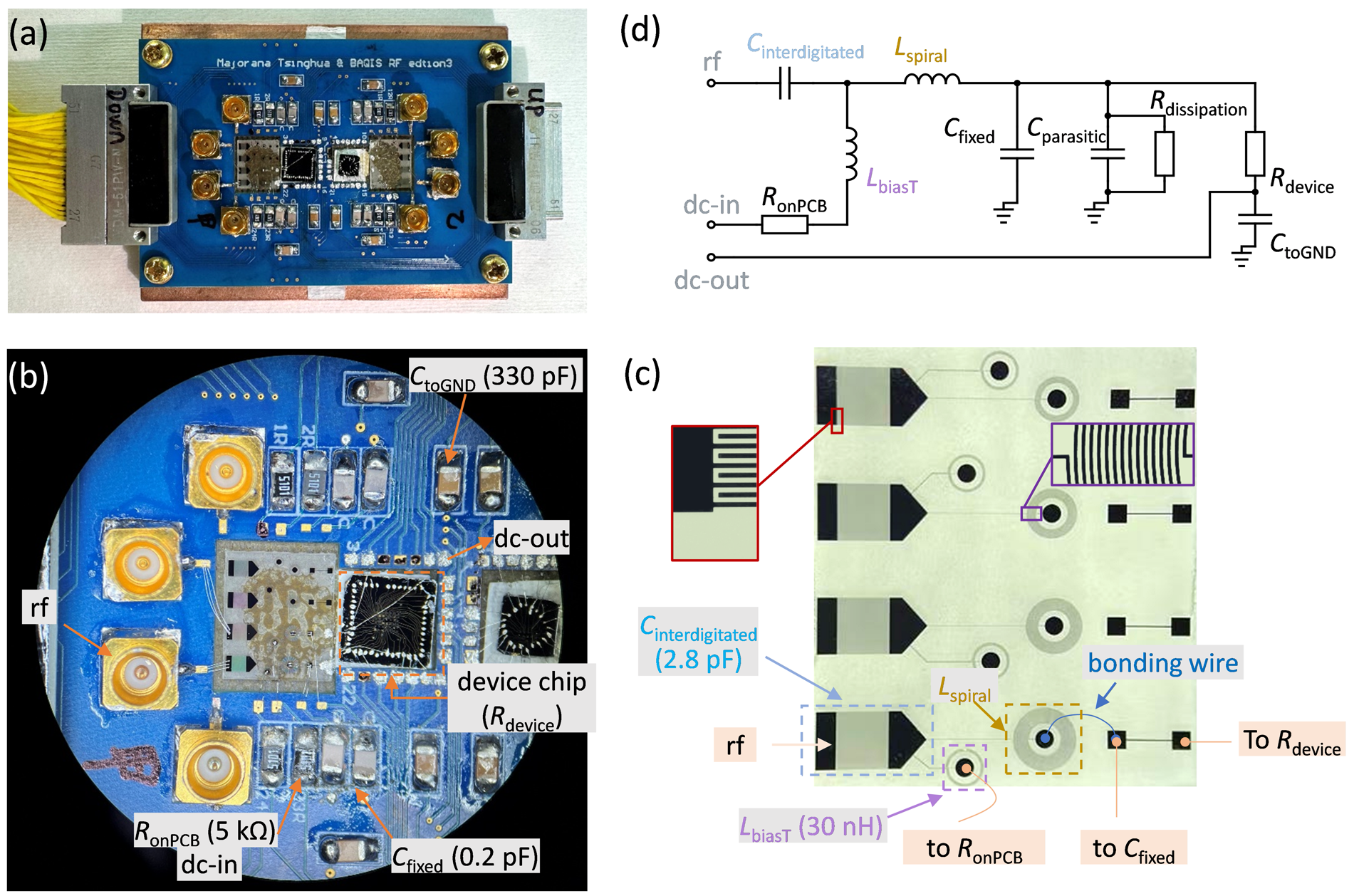}
\centering
\caption{(a) Optical image (full range) of the PCB.  (b) Zoom-in of the left part of the PCB. (c) Optical image of the sapphire substrate before gluing onto the PCB. (d) Circuit diagram, replotted from Fig. 1(b).  } 
\label{fig2}
\end{figure*}

Figure 6(a) shows a full image of the PCB mounted and screw-tightened on a copper block. Two device chips were loaded into the voids near the PCB center. Each chip was assigned with four rf ports (mini-SMP connectors). The NS device corresponds to the circuit on the left side of the PCB, shown in Fig. 6(b). The device chip is marked by the orange dashed box. The sapphire substrate carrying the interdigitated capacitors and spiral inductors is glued to the left side of the device chip. Figure 6(c) shows this substrate before mounting onto the PCB. The black regions are sputtered Nb superconducting film (100 nm thick), while the remaining regions are the sapphire substrate. For comparison, we replot the circuit diagram in Fig. 6(d) and label the corresponding elements in Figs. 6(b-c).  Insets in Fig. 6(c) are zoom-in images of the capacitor and the inductor.

\section{Circuit simulation}

Figure 7 shows the simulated $|S_{21}|$ for the NS-device circuit as the device conductance is varied between 0 and $2e^2/h$. We find that setting $R_{\text{dissipation}}$ to 4 k$\Omega$ aligns the variation of the resonance dip ($\sim$ 2 dB) with our result in Fig. 2(b). The parameters of the other circuit elements are labeled in Fig. 1(b).

\begin{figure}[htb]
\includegraphics[width=0.85\columnwidth]{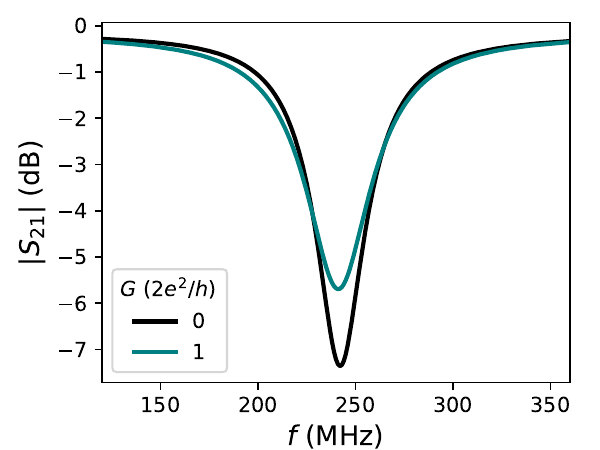}
\centering
\caption{$|S_{21}|$ simulation for $G$ = 0 (black) and $2e^2/h$ (green). $R_{\text{dissipation}} = 4$ k$\Omega$.} 
\end{figure}
%\FloatBarrier

\section{rf response to in-plane magnetic field}

\begin{figure*}[htb]
\includegraphics[width=\textwidth]{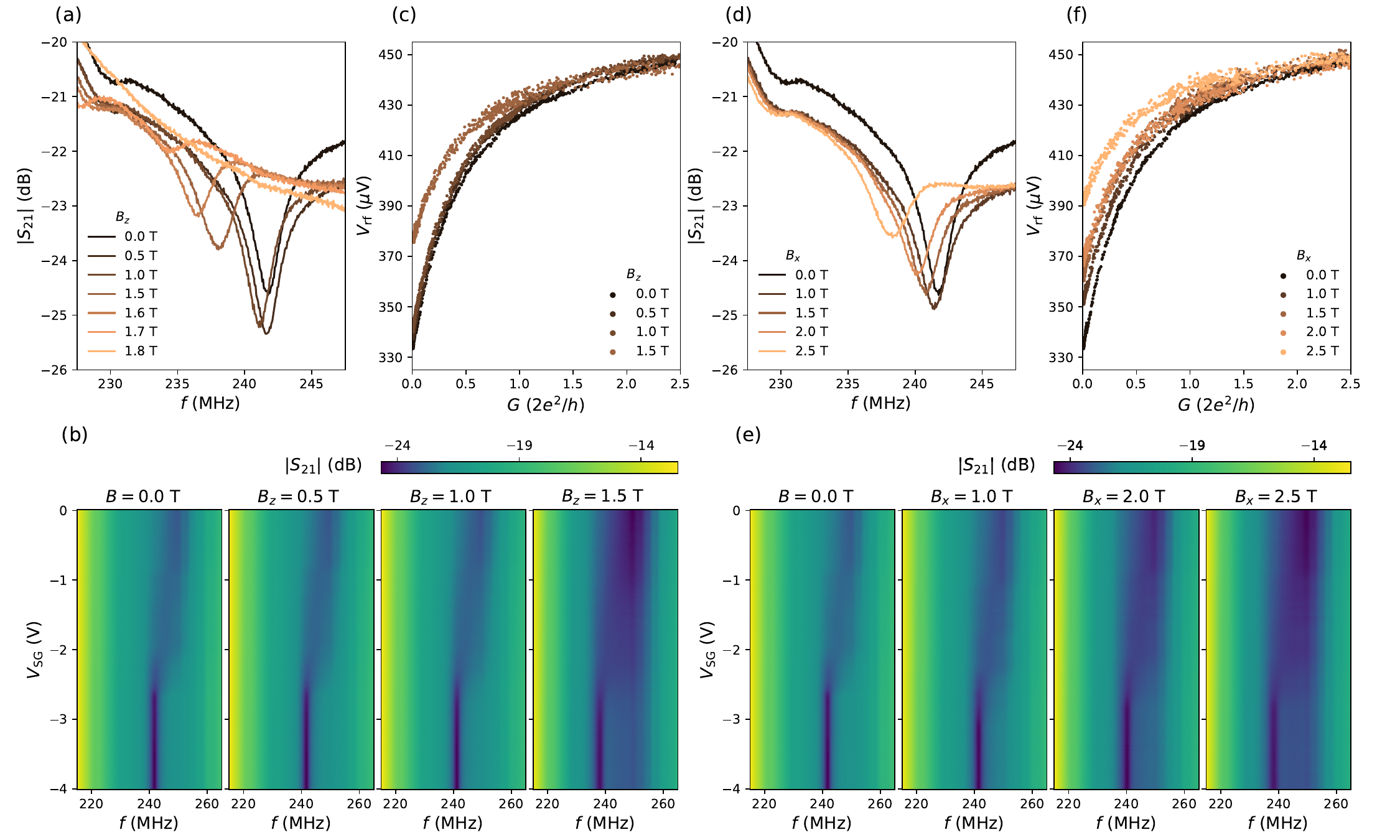}
\centering
\caption{(a-c) and (d-f) are similar to that in Fig. 4, but for $B_z$ and $B_x$, respectively.  $V_{\text{TG}} = -5$ V, $V = -2$ mV. $V_{\text{SG}} = -4$ V for (a) and (d). For (c) and (f), vertical offset was added to each curve so that $V_{\text{rf}}$ values at $G= 2.5 e^2/h$ are all aligned.} 
\end{figure*}

Figure 8 shows the cavity response for magnetic fields applied parallel to the device chip and the sapphire substrate. The spiral inductor can remain superconducting at larger in-plane fields due to the weaker orbital effect. Correspondingly, the resonance dip barely changes up to 1 T for both $B_z$ and $B_x$, as shown in Figs. 8(a) and 8(d). This behavior is also evident from the $V_{\text{rf}}$ versus $G$ mapping in Figs. 8(c) and 8(d). The resonance frequency decreases with increasing $B$.

\bibliography{mybibfile}% Produces the bibliography via BibTeX.

%\newpage

%\onecolumngrid

%\newpage
%\includepdf[pages=1]{PbTe_Reducing_Disorder_SM.pdf}

\end{document}